\documentclass[nofootinbib,onecolumn,showpacs,showkeys]{revtex4}
\usepackage{graphicx,color}
\usepackage{amssymb}
\usepackage{amsmath}
\usepackage{bm}
\usepackage[normalem]{ulem}
\usepackage{hyperref}

\begin{document}

\title{Nonconservative traceless type gravity}

\author{Mahamadou Daouda$^{1}$}\email{daoudah77@gmail.com}
\author{J. C. Fabris$^{2,4}$ }\email{fabris@pq.cnpq.br}
\author{A. M. Oliveira$^3$}\email{adriano.oliveira@ifes.edu.br}
\author{F. Smirnov$^5$}\email{fsmirnov94@gmail.com}
\author{H. E. S. Velten$^{6}$}\email{hermano.velten@ufop.edu.br}
\affiliation{$^1$Universit\'e Abdou Moumouni de Niamey, Niamey, Niger}
\affiliation{$^2$N\'ucleo Cosmo-ufes \& Departamento de F\'{\i}sica, CCE, Universidade Federal do Esp\'{\i}rito Santo (UFES), Vit\'oria, Brazil}
\affiliation{$^3$Instituto Federal do Esp\'irito Santo (IFES), Guarapari, Brazil}
\affiliation{$^4$National Research Nuclear University MEPhI, Kashirskoe sh. 31, Moscow 115409, Russia}
\affiliation{$^5$Saint Petersburg State University, St.Petersburg State University, 7/9 Universitetskaya nab., St. Petersburg, 199034 Russia}
\affiliation{$^6$Departamento de F\'isica, Universidade Federal de Ouro Preto, Campus Morro do Cruzeiro, 35400-000, Ouro Preto, Minas Gerais, Brazil}

\begin{abstract}
Extensions of the gravity theory in order to obtain traceless field equations have been widely considered in the literature. The leading example of such class of theories is the unimodular gravity, but there are other possibilities like
the mimetic gravity and the Rastall gravity with a coupling parameter $\lambda = 1/2$. The unimodular gravity proposal is a very interesting approach in other to address the cosmological constant problem. When coupled to matter such theories may imply that the energy-momentum tensor is not divergence free anymore. In this paper, a unimodular type theory will be developed by evading the conservation $T^{\mu\nu}_{\, ; \mu}=0$. The cosmological consequences of the later, both at background as well as for scalar and tensor perturbations, are explored. Possible further extensions of this approach are discussed as well as its connection with the traditional unimodular gravity. 
\end{abstract}

\maketitle

\section{Introduction}

Unimodular gravity can be seem as a simple modification (or gauge fixed version) of Einstein's general relativity. It has been realised long ago (see for example \cite{wei}) that by adding a constraint on the determinant of the metric to the action principle the same structure of the standard GR equipped with a cosmological constant term is achieved. Indeed, one can not argue that the cosmological constant problem is solved, but rather it appears as an initial condition which has to be set according to the cosmological evolution. Therefore, there is naturally great interest in analysing the relevance of unimodular gravity to the cosmological constant problem. In this work we will made use of the term unimodular gravity but having in mind actually a broader class of traceless gravitational theories.

It has been also widely investigated whether or not unimodular gravity is really indistinguishable from GR + $\Lambda$ concerning all possible physical aspects \cite{diego}. Although the background cosmological expansion is clearly the same in both cases it is not trivial that perturbations should follow the same structure. However, it has been show that the cosmological perturbations are identical (or one is not able to distinguish between them) in both theories \cite{Basak:2015swx, Gao:2014nia} under the condition that the usual conservation laws are also verified in the unimodular context. Apart from from the equivalence at the classical level, the equivalence at quantum level has been also discussed in Refs. \cite{Padilla:2014yea}.

There are also attempts in the literature to find out unimodular versions of extended theories of gravity as for example the unimodular $f(R)$ \cite{diego,Nojiri:2015sfd,Nojiri:2016ygo,Nojiri:2016plt}, unimodular $f(R,T)$ \cite{Rajabi:2017alf},  unimodular $f(\mathcal{G})$ \cite{Houndjo:2017jsj}. 
Another interesting similar proposal is the mimetic gravity proposed in Ref. \cite{mukhanov} shares with the unimodular gravity the property of traceless field equation \cite{Nojiri:2016ppu}. Another recent proposal for a traceless gravity in the inflationary context appears in Ref. \cite{Barrow:2019gup}. Also, the Rastall field equations \cite{Rastall:1973nw} have a similar structure to unimodular gravity when the
free parameter of the theory takes the value $\lambda = 1/2$ (see, for example \cite{Oliveira:2016ooo}). In these examples, there is no constraint on the determinant of the metric as in the case the traditional unimodular gravity. 

The goal of this work is to study the structure of unimodular gravity type theories by evading the imposition that the energy-momentum tensor has to be conserved. In this way, we have in mind the sort of extended unimodular gravity as cited above. Indeed, $T^{\mu\nu}_{\, ; \mu}=0$ has to be assumed in the derivation of the standard (or Einstein-Hilbert) version of the unimodular gravity. We promote this discussion in detail in the next sections. We are motivated by the phenomenology behind the Rastall theory with $\lambda = \frac{1}{2}$. We note, {\it en passant}, that there has been recently a discussion on the equivalence between Rastall and General Relativity theories \cite{visser,darabi}. However, as it will become clear later, our approach lies outside such discussion. 
 
We start Section II reviewing the standard unimodular theory. This allows us to show clearly at which point our proposal deviates from the standard case in which $T^{\mu\nu}_{\, ; \mu}=0$.

By giving up the usual conservation of the energy-momentum tensor, the general structure of the unimodular gravity, with the emergence of the cosmological constant, is not necessarily verified anymore. Moreover, some extra conditions must be imposed to the field equations in order to obtain a closed set of equations
\footnote{The necessity to impose some extra condition may be connected to the fact that a non-conservative theory behaves as an open system.}. We start exploring the specific extra condition of a constant Ricci scalar $R = cte$ which represents the simplest possible choice. This analysis is employed in Section III. This leads to a cosmological evolution exhibiting a transition from radiative phase to a de Sitter phase. We will show that combining background and perturbative analysis, the resulting model is more adapted to the primordial Universe. Moreover, by choosing a specific ansatz for the perturbed quantities the non conservative unimodular type model becomes, at least in the cosmological case, identical to the corresponding GR case, except for some multifluid configurations. In Section IV we investigate another ansatz in which the combination $\sqrt{-g}\{R + 4\Lambda\}$ is constant. Then, by relaxing the condition $R = cte$ at background level many interesting new features appear. Indeed, it is verified that the $R = cte$ hypothesis represents only a particular case of the general formalism presented in Section IV. We conclude in section V.

\section{Non-Conservative version of unimodular gravity}

\subsection{A short review of ``standard" unimodular gravity}

Let us consider the equations of the unimodular gravity written in the following form
\begin{eqnarray}\label{fe2}
G_{\mu\nu} + \frac{1}{4}g_{\mu\nu}R = 8\pi G\biggr\{T_{\mu\nu} - \frac{1}{4}g_{\mu\nu}T\biggl\}.
\end{eqnarray}
This equation can be obtained from the Einstein-Hilbert Lagrangian by imposing the condition $g_{\mu\nu}\delta g^{\mu\nu} = 0$ on the variation of the metric \cite{diego}.

Taking the divergence of this equation, and using the Bianchi identities, we find,
\begin{eqnarray}
\label{cons}
\frac{R^{,\nu}}{4} = 8\pi G\biggr\{T^{\mu\nu}_{;\mu} - \frac{T^{,\nu}}{4}\biggl\}.
\end{eqnarray}

It is widely accepted that the unimodular gravity comes from an action, which is invariant by diffeomorphism. Hence, the energy-momentum tensor conserves.  However, this issue is a crucial and subtle aspect of the unimodular gravity. Since one has now an the extra constraint on the determinant of
the metric the theory is left with 9 independent component equations of motion, one less than General Relativity. This impedes the use of the Bianchi identities for deriving the conservation of the energy-momentum tensor. This viewpoint is also adopted in Ref. \cite{Gao:2014nia}. Therefore, one has indeed to impose the conservation
\begin{equation}\label{usualcons}
T^{\mu\nu}_{\, ;\mu}=0.
\end{equation}

Then, by imposing the use of the above equation we are left with the following result for the standard unimodular gravity,
\begin{eqnarray}\label{cons2}
 R^{,\nu} = - 8\pi GT^{,\nu}.
\end{eqnarray}
The above equation can now be integrated leading to an expression for the Ricci scalar in terms of the trace of the energy-momentum
\begin{eqnarray}
R = - 8\pi G T + 4\Lambda,
\end{eqnarray}
where $\Lambda$ is an integration constant. Inserting this result in the field equations (\ref{fe2}) we find,
\begin{eqnarray}
G_{\mu\nu} = 8\pi GT_{\mu\nu} + \Lambda g_{\mu\nu}.
\end{eqnarray}
Hence, unimodular gravity is identical to General Relativity with a cosmological constant which appears as an integration constant.

\subsection{Avoiding the conservation of $ T^{\mu\nu}$ and cosmological implications}

According to Ref. \cite{Alvarez:2007nn} the integrability condition applied to unimodular gravity should be rather $( T^{\mu\nu} / \sqrt{\left|g\right|})_{\,\, ;\mu}=0$ but, by quoting this reference {\it this is an extra condition that should be added for consistency but one that looks a bit mysterious}. Recently, is has been found in Ref. \cite{Barcelo:2014mua} that by allowing the self-coupling of the spin-2 field Eq. \ref{usualcons} emerges naturally as a consequence of Poincar\'e invariance of the theory.

In the approach for unimodular gravity we are considering, the conservation law is an {\it ad hoc} hypothesis. Other approaches are possible, see \cite{wei,diego}, but we adopt here the one presented in Ref. \cite{Gao:2014nia}. Our main goal here is to evade the hypothesis leading to (\ref{cons}). This implies the necessity to impose other {\it ad hoc} hypothesis, of course. In doing so, the features of the unimodular usual hypothesis can be tested and possible variants further analysed.

In order to evade $T^{\mu\nu}_{;\mu}=0$ and to design a non-conservative version of the unimodular theory we use the Rastall theory as inspiration. In Rastall gravity, there is no action. Consequently, the divergence of the energy-momentum tensor is not necessarily zero, and the integral of the
conservation law is not possible as it occurs with unimodular gravity. We remain with equation (\ref{cons}) in its complete form. Then, in the approach we are about to develop we assume we can not promote the step from Eq.(\ref{cons}) to Eq.(\ref{cons2}). That is all difference we must exploit. Hence, our fundamental structure is given for the moment by equation (\ref{cons}).

Let us consider the flat cosmological case equipped with a flat Friedmann-Robertson-Lemaitre-Walker (FLRW) metric
\begin{equation}
ds^{2}= dt^2-a^2(t) (dx^i)^2.
\end{equation}
According to the above metric the components of the Ricci tensor are
\begin{eqnarray}
R_{00} &=& - 3\dot H - 3 H^ 2,\\
R_{ij} &=& (\dot H + 3 H^2)a^2\delta_{ij},
\end{eqnarray}
and the Ricci scalar is,
\begin{eqnarray}
R = - 6(\dot H + 2 H^2).
\label{Rscalar}
\end{eqnarray}
The energy-momentum tensor is given by,
\begin{eqnarray}
T_{\mu\nu} = (\rho + p)u_\mu u_\nu - p g_{\mu\nu},
\end{eqnarray}
where its components and its trace read
\begin{eqnarray}
T_{00} &=& \rho,\\
T_{ij} &=& p a^2\delta_{ij},\\
T &=& \rho - 3 p.
\end{eqnarray}

It is worth noting that the $0-0$ and $i-j$ components of the fundamental unimodular equations (\ref{fe2}) lead to the same result
\begin{eqnarray}
\label{febis}
\dot H = - 4\pi G(\rho + p).
\end{eqnarray}
The implications of this structure for a primordial inflation based on a cosmological constant or a self-interacting scalar field have been discussed in Refs. \cite{ellis,barrow} where some interesting analysis of a trace-free theory of gravity are presented.

The equation (\ref{cons}) is just a combination of the fundamental unimodular equation (\ref{fe2}) and its derivative. Thus, it does not contain in principle any additional information. In the standard unimodular gravity this does not represent a crucial issue since the conservation of the energy-momentum tensor is satisfied, leading in the cosmological case to
\begin{eqnarray}
\dot\rho + 3H(\rho + 3p) = 0.
\label{cont}
\end{eqnarray}
With the specification of the equation of state parameter $w=p/ \rho$ the system of equations is closed.
However, in our alternative approach, the conservation of the total energy-momentum tensor does not exist and extra assumptions have to be imposed in order to have a complete set of equations.

An important remark must be made at this point. In the unimodular theory, the condition $\sqrt{-g} = $ constant is a fundamental property that leads to the General Relativity equations equipped with $\Lambda$ as an integration constant. In reference \cite{wei} it is argued that this condition allows to recover the general covariance by diffeomorphism. Here, on the other hand we have the traceless version of the gravitational equations
and they are our fundamental set up. A traceless set of equations in the geometric gravitational context does not close the necessary number of equations in order to integrate them unambiguously, at least in the cosmological framework. This has been discussed briefly in the Ref. \cite{Oliveira:2016ooo}, and a possible way out was sketched. Such situation depends on the symmetry of the problem: using static spherically symmetric configurations such necessity does not appear, as it has also been discussed in Ref. \cite{Oliveira:2016ooo}. Such properties have some similarities with conformal theories of gravity \cite{man}, what is somehow expected since the fundamental equations are
traceless.

Here we extend the considerations sketched in Ref. \cite{Oliveira:2016ooo} for the cosmological case, both at background and perturbative levels. In order to do this, finding a specific cosmological scenario, we must impose an suitable additional condition to close the set of equations. But this condition has a different status in comparison with the unimodular one: in the present approach we look for cosmological solutions subjected to specific restrictions that otherwise would remain underdetermined. These restrictions do not appear as a fundamental aspect of the theory as in the unimodular case since it depends on the symmetry of the problem as already stated. This fact allows to freely choose the coordinate system we will work. This is not the case in the usual unimodular context \cite{Gao:2014nia}. As we will see, using such approach the main aspects of a corresponding scenarios in GR may be recovered at background level, but with possible different properties at perturbative level.

\section{Theory constrained by a constant Ricci scalar}

The formalism discussed above lacks a specific choice to constrain the Ricci scalar which holds the remaining degree of freedom of this formalism. Among all the possibilities the simplest one is to consider a constant curvature 
\begin{equation}
R = cte.
\label{Rcte}
\end{equation}
With this hypothesis Eq. (\ref{Rscalar}) becomes
\begin{eqnarray}
\dot H + 2H^2 = k,
\end{eqnarray}
being $k$ a constant factor. For $k > 0$, one finds the usual de Sitter solution
\begin{eqnarray}
a = a_0e^{\sqrt{\frac{k}{2}}t}.
\end{eqnarray}
It is worth noting that this is a solution for GR with a cosmological constant also.
Moreover, there are two disconnected branches represented by the following solutions.
\begin{enumerate}
\item First branch (non-singular solution):
\begin{eqnarray}
a = a_0 \cosh^\frac{1}{2}\sqrt{2k}t,\;
\end{eqnarray}
\item Second branch (singular solution):
\begin{eqnarray}
a = a_0 \sinh^\frac{1}{2}\sqrt{2k}t.
\end{eqnarray}
\end{enumerate}
The first branch, the non-singular solution, implies in a violation of the null energy condition, while the second branch, the singular solution, is achieved when the null energy condition is satisfied.
In principle, these two branch solutions can also be achieved in the GR context using radiation and a cosmological constant sourcing the field equations. But, remark that, in our model, the only condition imposed in order to obtain
those solutions is $\rho \neq - p$. Besides this condition, no equation of state has to be specified. In fact, the radiative-like feature appears from
the traceless property of the field equations, while the cosmological constant emerges naturally as an integration constant.

For $k < 0$, the solution is
\begin{eqnarray}
a = a_0 \sin^\frac{1}{2}\sqrt{2|k|}t.
\end{eqnarray}
The second branch, represented by the cosines solution, is just a spatial displacement of the sines solution.

For $k = 0$, we have the typical radiative solution,
\begin{eqnarray}
a = a_0 t^\frac{1}{2}.
\end{eqnarray}

Now, with this hypothesis of constant curvature, we can investigate two situations concerning the matter sector.
Remember that, since $R =$ constant, the equation (\ref{cons}) for the matter sector becomes
\begin{eqnarray}
\label{cons-bis}
{T^{\mu\nu}}_{;\mu} = \frac{T^{;\nu}}{4}.
\end{eqnarray}

\begin{itemize}
\item Single fluid case.

From equation (\ref{cons-bis}), we obtain the following result
\begin{eqnarray}
\dot\rho + \dot p + 4H(\rho + p ) = 0,
\end{eqnarray}
with the solution for the enthalpy of the system
\begin{eqnarray}
\label{enth}
\rho + p = Ca^{-4},
\end{eqnarray}
$C$ being a constant. The relevant r\^ole played here by the enthalpy of the system can be connected to the emergent gravity proposal \cite{pad}, but at the present stage this is pure speculative possible connection. The solution (\ref{enth}) corresponds to the typical radiative behavior. Remark also that equations are sensitive to the enthalpy $\rho + p$ only, independent on the equation of state adopted. 

\item Double fluid case.

Now, the solution for the scale factor, under the hypothesis of constant curvature is the same as before.
Let us suppose however that the energy-momentum tensor is given by,
\begin{eqnarray}
T^{\mu\nu} = T^{\mu\nu}_m + T^{\mu\nu}_x,
\end{eqnarray}
where $m$ stands for matter, and $x$ indicates a generic exotic fluid.
Let us suppose also that the matter component is pressureless and does conserve separately.

The matter field behaves as,
\begin{eqnarray}
\rho_m = \frac{\rho_m}{a^3},
\end{eqnarray}
while equation (\ref{cons-bis}) implies for the exotic fluid,
\begin{eqnarray}
\dot\rho_x + \dot p_x  +  4(\rho_x + p_x) = \frac{1}{3}\dot\rho_m,
\end{eqnarray}
resulting in the solution,
\begin{eqnarray}
\rho_x + p_x = \frac{\rho_0}{a^4} - \frac{\rho_{m0}}{a^3}.
\end{eqnarray}
Some interesting features of this solution:
\begin{enumerate}
\item Again, for the fluid $x$, only the combination $\rho_x + p_x$ is relevant;
\item In the past, the fluid $x$ behaves as a radiative fluid and drives the dynamical evolution of the Universe;
\item As matter begin to dominate the energy content of the Universe, the fluid $x$ begins to behave repulsively;
\item The case $p_x = - p_x$ (cosmological constant) only the empty Universe is possible, and we recover the unimodular Universe
with a cosmological constant emerging as an integration constant.
\end{enumerate}
\end{itemize}

In the previous cases
\begin{eqnarray}
\dot H = - \frac{A}{a^4},
\end{eqnarray}
where $A$ is a positive constant. Hence, the super inflationary phase is a characteristic of this model. Initially, for $k > 0$ the Universe exhibit a typical radiative behavior reaching an asymptotic de Sitter phase. In principle, a matter dominated era may occur meanwhile, but it must be verified if it last enough for the formation of structure to take place. We will verify later that this is not the case, at least under the particular conditions assumed above.

\subsection{Behavior of the new background solution}

Let us consider the background solution represented by,
\begin{eqnarray}
a = a_0\sinh^\frac{1}{2}\kappa t,
\label{aa0}
\end{eqnarray}
where $\kappa=\sqrt{2k}$. We normalise the scale factor such that today, $t = t_0$, it is equal to unity. Hence,
\begin{eqnarray}
a_0 = \frac{1}{\sinh^\frac{1}{2}\kappa t}.
\end{eqnarray}

We write the fundamental equation as,
\begin{eqnarray}
\dot H = - 4\pi G(\rho + p) = - 4\pi G\tilde \rho = - 4\pi G\frac{\tilde \rho_0}{a^4}.
\end{eqnarray}
With above solution (\ref{aa0}) the above equation can be integrated leading to
\begin{eqnarray}
H = \frac{4\pi G\tilde\rho_0}{\kappa a_0^4}\coth \kappa t + c,
\end{eqnarray}
where $c$ is an integration constant. For simplicity and without loss of generality we consider $c = 0$. In any case, it can be absorbed in the definition of the expansion rate $H$.

The solution for $H$ can then be rewritten as,
\begin{eqnarray}
H = \frac{4\pi G\tilde \rho_0}{\kappa a_0^4}\sqrt{\frac{(1 + z)^4}{a_0^4} + 1}
\end{eqnarray}
where the redshift hsa been used as dynamical variable $1+z=a_0 /a$. The constants may be fixed using the today's Hubble constant parameter $H(z = 0)=H_0$ and also making use of the fact that at the asymptotic future limit given by $z = - 1$ a de Sitter phased is reached such that $H = \Lambda$.
With such conditions, the expression for the Hubble function reduces to,
\begin{eqnarray}
H = H_0\sqrt{(1 - \tilde \Lambda^2)(1 + z)^4 + \tilde \Lambda^2},
\label{H}\end{eqnarray}
where
\begin{eqnarray}
\tilde \Lambda = \frac{\Lambda}{H_0}.
\end{eqnarray}
This solution has two free parameters, $H_0$ and $\tilde{\Lambda}$, the same number of degrees of freedom as in the flat $\Lambda$CDM model. It is worth noting that the model studied here given by Eq. (\ref{H}) is equivalent to a GR based expansion composed of radiation and cosmological constant. Although not suitable for the late time Universe expansion, where the presence of a matter component is mandatory, the solution given by Eq.(\ref{H}) can model the early Universe.
 
\subsection{Perturbations}

Since we have established in the previous section the background dynamics of the non-conservative unimodular cosmology, let us consider now the evolution of perturbations in the model described previously.  In principle, we could use any of the background solutions found before. We will analyse in detail 
the singular solution represented by the hyperbolic sines showing that even at perturbative level it is equivalent to the corresponding case in General Relativity, with radiation and cosmological constant. Later, we will discuss some interesting situations where the equivalence with GR is broken. 
We will not discuss the non singular solution represented by the hyperbolic cosines since it contains negative energy.

We start by splitting the metric such that
\begin{eqnarray}
g_{\mu\nu} = g^b_{\mu\nu} + h_{\mu\nu},
\end{eqnarray}
where $g^b_{\mu\nu}$ is the background metric and $h_{\mu\nu}$ represents a small fluctuation around it.
We develop the perturbative analysis by fixing synchronous coordinate conditions (or the synchronous gauge)
\begin{eqnarray}
h_{\mu0} = 0.
\end{eqnarray}
Consequently, the perturbed components of the Ricci tensor are \cite{weinberg}
\begin{eqnarray}
\delta R_{00} &=& \frac{\ddot h}{2} + H\dot h,\\
\delta R_{0i} &=& \frac{1}{2a^2}\biggr\{\dot h_{kk,i} - \dot h_{ki,k} - 2H\biggr( h_{kk,i} - h_{ki,k}\biggl)\biggl\},\\
\delta R_{ij} &=& \frac{1}{2a^2}\biggr\{\nabla^2 h_{ij} - h_{ik,j,k} - h_{jk,i,k} + h_{kk,i,j}\biggl\} \nonumber\\
&-& \frac{1}{2}\ddot h_{ij} + \frac{H}{2}\dot h_{ij} - \delta_{ij}Ha^2\frac{\dot h}{2} - 2H^2h_{ij}.
\end{eqnarray}
In the above expressions we have defined the synchronous scalar potential as usual
\begin{eqnarray}
h = \frac{h_{kk}}{a^2}.
\end{eqnarray}

The perturbation of the Ricci scalar is given by,
\begin{eqnarray}
\delta R = \ddot h + 4H\dot h - \frac{1}{a^2}\biggr\{\nabla^2 h - \frac{h_{kl,k,l}}{a^2}\biggl\}.
\label{deltaR}\end{eqnarray}

In order to obtain the full set of perturbed equation we also have to consider fluctuations of the field source of the theory. We adopt a perfect fluid structure
\begin{eqnarray}
T_{\mu\nu} = (\rho + p)u_\mu u_\nu - p g_{\mu\nu},
\end{eqnarray}
which has the following first order components
\begin{eqnarray}
\delta T_{00} &=& \delta\rho, \\
\delta T_{0i} &=& (\rho + p)\delta u_i,\\
\delta T_{ij} &=& a^2 \delta p \delta_{ij} - p h_{ij},\\
\delta T &=& \delta\rho - 3\delta p.
\end{eqnarray}
Thus the perturbed version of the field equations (\ref{fe2}) becomes
\begin{eqnarray}
\label{curv}
\delta R_{\mu\nu} - \frac{1}{4}(h_{\mu\nu}R + g_{\mu\nu}\delta R) = 8\pi G\biggr\{\delta T_{\mu\nu} - \frac{1}{4}(h_{\mu\nu}T + g_{\mu\nu}\delta T\biggl)\biggl\}.
\end{eqnarray}
From the above equation we can write down their components. They read
\begin{eqnarray}
\label{pe1}
\ddot h + 2H\dot h - \frac{\delta R}{2} &=& 12\pi G(\delta \rho + \delta p),\\
\label{pe2}
\frac{1}{2a^2}\biggr\{\dot h_{kk,i} - \dot h_{ki,k} - 2H\biggr( h_{kk,i} - h_{ki,k}\biggl)\biggl\} &=& 8\pi G(\rho + p)\delta u_i,\\
\frac{1}{2a^2}\biggr\{\nabla^2 h_{ij} - h_{ik,j,k} - h_{jk,i,k} + h_{kk,i,j}\biggl\} &-& \frac{1}{2}\ddot h_{ij} + \frac{H}{2}\dot h_{ij} - \delta_{ij}Ha^2\frac{\dot h}{2} + (\dot H + H^2)h_{ij}\nonumber\\
+ a^2\frac{\delta R}{4}\delta_{ij} &=& 2\pi Ga^2(\delta \rho + \delta p)\delta_{ij}.
\label{pe3}
\end{eqnarray}
It is worth noting that also at the perturbative level the enthalpy $\rho + p$ is the relevant quantity that obeys the first order dynamics. Equation (\ref{pe2}) is actually the expression for the perturbation of the four-velocity. 

Equation (\ref{pe3}) can be decomposed into two: its trace and its double divergence. The trace of (\ref{pe3}) reads 
\begin{eqnarray}
\label{pe3a}
\frac{1}{a^2}(\nabla^2h - g) - \frac{\ddot h}{2} - 3H \dot h + \frac{3}{4}\delta R = 6\pi G(\delta\rho + \delta p),
\end{eqnarray}
with the definition,
\begin{eqnarray}
g = \frac{h_{kl,k,l}}{a^2}.
\end{eqnarray}
Combining (\ref{pe3a}) with (\ref{pe1}) we find again the expression for $\delta R$, Eq. (\ref{curv}), confirming that the entire set of equations is self-consistent.

The divergence of (\ref{pe3}) combined with (\ref{pe1}) and the expression for $\delta R$ provides the following equation
\begin{eqnarray}
\frac{\nabla^2}{a^2}(\nabla^2 h - g) - 3(\ddot g + 3H\dot g) + \nabla^2\ddot h  + 3H \nabla^2\dot h = 0.
\end{eqnarray}

There are then two equations for three unknown functions namely, $\delta\tilde\rho = \delta\rho + \delta p$, $h$ and $g$.
As in the background case we have to impose an additional condition. According to the procedure adopted for the background
\begin{eqnarray}
\delta R = 0.
\end{eqnarray}
Applying the above ansatz to Eq. (\ref{deltaR}) we obtain,
\begin{eqnarray}
g = - a^2\biggr\{\ddot h + 4H\dot h - \frac{\nabla^2 h}{a^2}\biggl\}.
\label{g}\end{eqnarray}
By taking the derivative of this relation and using (\ref{pe1}), we obtain a single equation for $h$,
\begin{eqnarray}
3\biggr\{h^{iv} + 11 H\stackrel{\cdot\cdot \cdot}{h} + (10\dot H + 38H^2)\ddot h + 20H(\dot H + 2H^2)\dot h\biggl\}- \frac{\nabla^2}{a^2}(\ddot h + 2H\dot h) = 0.
\end{eqnarray}
By defining $F = \dot h$, we have a third order differential equation,
\begin{eqnarray}
3\biggr\{\stackrel{\cdot\cdot\cdot}{F} + 11 H\ddot F + (10\dot H + 38H^2)\dot F + 20H(\dot H + 2H^2)F\biggl\}- \frac{\nabla^2}{a^2}(\dot F + 2HF) = 0.
\end{eqnarray}

Replacing the time derivative $(``\,\,^{.}\,\,")$ by the derivative with respect to the scale factor $(``\,\,^{\prime}\,\,")$ the above equation reads
\begin{eqnarray}
F'''+ \biggr\{3\frac{H'}{H}  + \frac{14}{a}\biggl\}F'' + \biggr\{20\frac{H'}{aH} + \frac{50}{a^2}\biggl\}F' + 20\biggr\{\frac{H'}{a^2H} + \frac{2}{a^3}\biggl\}F
= \frac{\nabla^2}{3a^4H^2}\biggr(F' + 2\frac{F}{a}\biggl).
\label{F}\end{eqnarray}


The synchronous coordinate condition does not fix completely the coordinate system. There is a residual coordinate freedom which leads to a spurious mode represented by,
\begin{eqnarray}
F_g = \dot h_g = - 3\dot H\Psi(\vec x) - \frac{\nabla^2\Psi(\vec x)}{a^2},
\end{eqnarray}
where the subscript $g$ indicates the functional form of the residual gauge mode, and $\Psi$ is an arbitrary function of the position.
It is interesting that the condition $\delta R = 0$ is still satisfied for the gauge mode.

Using this gauge mode we can lower the order of the perturbed differential equation. Then it reads now
\begin{eqnarray}
\label{spe}
\lambda'' + \left\{ 3\frac{F'_g}{F_g} + 3\frac{H'}{H} + \frac{14}{a} \right\}\lambda' + \left\{ 3\frac{F''_g}{F_g} + 2 \frac{F'_g}{F_g}\left( 3\frac{H'}{H} + \frac{14}{a} \right) + \frac{20}{a} \frac{H'}{H} + \frac{50}{a^2} \right\} \lambda = - \frac{k^2}{3 a^4 H^2} \lambda,
\end{eqnarray}
where we have defined $F = F_g \int \lambda da$ and a Fourier decomposition has been made.

Even with this simplification the equation is still too complicated to be solved analytically. However, it can be solved asymptotically.
In the limit $a \rightarrow \infty$, the Hubble function becomes $H = \Lambda$, a constant. The equation simplifies to,
\begin{eqnarray}
\lambda'' + 8\frac{\lambda'}{a} + 12\frac{\lambda}{a^2} = 0,
\end{eqnarray}
where we have discarded the right hand side of equation (\ref{spe}), since it is, in that limit, exponentially suppressed.
This implies a solution under the form,
\begin{eqnarray}
F \propto e^{p_\pm\Lambda t}, \quad p_+ = - 5, \quad p_- = -6.
\end{eqnarray}
Hence, the perturbations are exponentially suppressed as expected since this limit corresponds to the de Sitter regime.

\begin{figure}[t]
\includegraphics[width=0.5\textwidth]{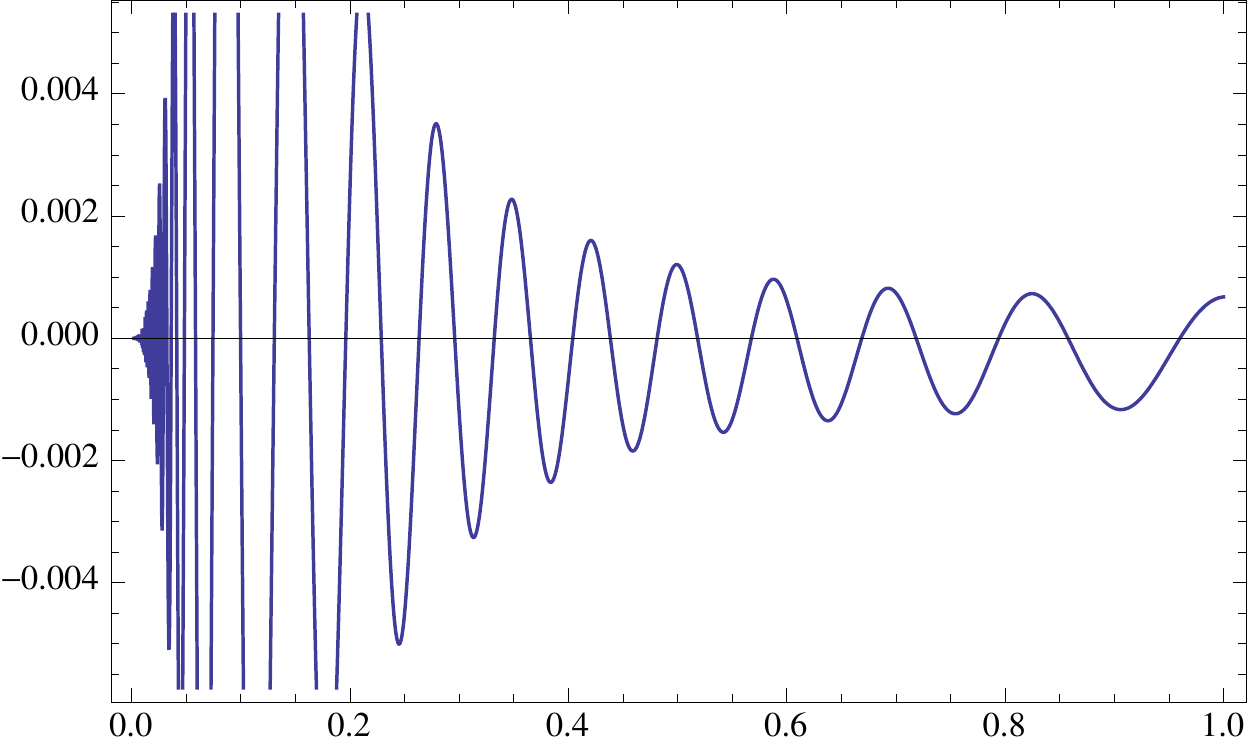}
\caption{\label{fig:i} Behavior for the perturbed quantity $F$ as function of of the scale factor, for $k = 0.01$ and $\tilde\Lambda = 0.9$.}
\label{pert}
\end{figure}

In the other limit, $a \rightarrow 0$, the background regime lies in the radiative phase, i.e., the Hubble function reads $H = \frac{\Omega}{a^2}$. In this case, using the knowledge of the gauge mode and redefining $\lambda = a^\frac{5}{2}\gamma$, we obtain the equation,
\begin{eqnarray}
\gamma'' + \frac{\gamma'}{a} + \biggr\{\tilde k^2 - \frac{1}{4a^2}\biggl\}\gamma = 0,
\end{eqnarray}
where $\tilde k = k/\sqrt{3}$.
The above equation is a Bessel differential equation with solution
\begin{eqnarray}
F = \frac{c_\pm}{a^4}\int a^\frac{5}{2}J_{\pm\frac{1}{2}}(\tilde ka)da.
\end{eqnarray}
This solution implies oscillations in the perturbed quantities. Thus, the amplitude of the perturbed scalar quantities oscillate initially and suffer later a exponential suppression. Indeed this behavior is the expected one for a Universe that behaves initially as radiation dominated and reachs a de Sitter phase. Such behavior deduced from the asymptotic expressions is confirmed by numerical integration of the full equation (\ref{F}) as shown in Fig. \ref{pert}.

The model features found above are also present in the GR context with a cosmological constant and radiation. This equivalence is consequence of the
ansatz $\delta R = 0$. However, in principle this implies to convert the gauge mode into the physical mode unless another gauge invariant ansatz is imposed. 

A particular case occurs if the long wavelength limit is taken. Hence, spatial derivatives can be neglected and $\nabla^2 h = g = 0$. Then Eq. (\ref{g}) becomes
\begin{eqnarray}
\ddot h + 4 H\dot h = 0,
\end{eqnarray}
with solution,
\begin{eqnarray}
h \propto \coth(\sqrt{2k}t),
\end{eqnarray}
which is essentially the same solution as in the $\Lambda$CDM case.

Indeed, once the solution for $h$ is found the behavior of matter perturbations is given by (\ref{pe1}). However, there is an interesting situation when
the energy-momentum tensor is decomposed into two components, as described above, an ordinary matter and an extra component, a situation that can not be fitted in the GR context (if the matter component conserves, the second component also conserves necessarily, and the solution can not be given by the hyperbolic sinus anymore).

Since the pressureless matter component conserves separately
\begin{eqnarray}
\delta_m = h.
\end{eqnarray}
In the long wavelength limit, this implies that the matter component behaves as in the $\Lambda$CDM model. The difference comes from the modes inside the horizon: in the $\Lambda$CDM model the matter component follows the same behavior for all wavelength, while for our case the behavior inside the horizon follows a typical pattern of the radiative perturbations.

It is worth noting that we end up this section showing that condition $R=cte$ imposed to the non-conservative version of the unimodular gravity leads to the same solutions as standard GR. We explore in the next section another constraining relation involving the Ricci scalar in which the equivalence is broken.

\section{ A possible nontrivial extension}

In the previous analysis, in which we adopted a constant Ricci scalar we ended up with a radiative fluid in presence of a cosmological constant. Matter may be introduced by splitting the energy momentum tensor into two components. However, we have another possibility to explore in order to obtain a full matter sector, with radiation, pressureless matter and a cosmological term. Nevertheless, the model of the last section resembles the same dynamics  of GR.

Now, we explore another condition imposed to the Ricci scalar which lead to different features when compared to GR. Let us consider now the following hypothesis:
\begin{eqnarray}
\sqrt{-g}\{R + 4\Lambda\} = l,
\label{R4Lambdak}
\end{eqnarray}
where $\Lambda$ and $l$ are constants. This hypothesis, together with the unimodular condition on the determinant of the metric, implies in supposing that the usual Einstein-Hilbert Lagrangian is a fixed point.

The choice (\ref{R4Lambdak}) represents a general case within the non-conservative unimodular approach studied in this work. The case studied in the section above (\ref{Rcte}) corresponds to $l=0$ in (\ref{R4Lambdak}).

With this new hypothesis, the field equations take the form,
\begin{eqnarray}
\label{ne1}
R_{\mu\nu} - \frac{1}{2}g_{\mu\nu}R &=& 8\pi G\biggr(T_{\mu\nu} - \frac{1}{4}g_{\mu\nu}T\biggl) + g_{\mu\nu}\Lambda - \frac{l}{4}\frac{g_{\mu\nu}}{\sqrt{-g}},\\
{T^{\mu\nu}}_{;\mu} - \frac{1}{4}T^{;\nu} &=& 0.
\label{Rmunu2}
\end{eqnarray}
Of course, such equations can not be invariant by diffemorphism due to the presence of tensorial densities in the last term of
(\ref{ne1}). However, they are invariant by a restricted class of transformations one clear example being that dictated by the unimodular condition.

In order to verify the consistency of the equation (\ref{R4Lambdak}) it is important to remember that the derivative of a tensor density
$\sqrt{-g}$ is given by,
\begin{eqnarray}
\sqrt{-g}_{;\mu} = \sqrt{-g}_{,\mu} - \Gamma^\rho_{\mu\rho}\sqrt{-g}.
\end{eqnarray}
We can rewrite  Eq. (\ref{Rmunu2}) by defining 
\begin{eqnarray}
\tilde T^{\mu\nu} = T_{\mu\nu} - \frac{1}{4}g_{\mu\nu}T.
\end{eqnarray}
Equation (\ref{Rmunu2}) resembles the GR equations sourced by a cosmological term and - as it will be seen later - a geometric matter (dark matter) component.

Using the FLRW metric, the resulting equations are
\begin{eqnarray}
&\dot H&= - 4\pi G(\rho + p),\\
&\dot\rho&+ \dot p + 4H(\rho + p) = 0.
\end{eqnarray}
The last equation has the solution (writing the constant in a convenient way),
\begin{eqnarray}
\rho + p = \frac{\rho_{\gamma0}}{a^4}.
\end{eqnarray}
Using the hypothesis for $R$ (\ref{R4Lambdak}) with the expression (\ref{Rscalar}) the final {\it Friedmann} equation is,
\begin{eqnarray}
3H^2 = 8\pi G\frac{\rho_{\gamma0}}{a^4} - \frac{l}{4a^3} + \Lambda.
\label{Ha3}
\end{eqnarray}
It reduces effectively to the $\Lambda$CDM corresponding equation if,
\begin{eqnarray}
l = - 32\pi G\rho_{m0}.
\end{eqnarray}

Hence, it is worth noting that the flat $\Lambda$CDM background expansion is achieved
\begin{eqnarray}
3H^2 = 8\pi G\biggr(\frac{\rho_{\gamma0}}{a^4} + \frac{\rho_{m0}}{a^3}\biggl) + \Lambda.
\end{eqnarray}
Notice that matter like component appearing in (\ref{Ha3}) has a geometric origin. Baryons, as it will be discussed latter, can be recovered from a decomposition of the radiative-like fluid.

\subsection{Perturbations}

Our goal now is to find the perturbed equations considering the new constraint \ref{R4Lambdak}. Using the formalism of the synchronous gauge as presented in the last section we obtain the $0-0$ component of the field equation identically to Eq. (\ref{pe1}).

Differently now, we must perturb the condition (\ref{R4Lambdak}). Using the fact that (remember that if $h_{\mu\nu} = \delta g_{\mu\nu}$, $\delta g^{\mu\nu} = - h^{\mu\nu}$),
\begin{eqnarray}
\delta\sqrt{-g} =  - \frac{1}{2}\sqrt{-g}g_{\mu\nu}\delta g^{\mu\nu} =- \frac{1}{2}\sqrt{-g} h,
\end{eqnarray}
we obtain,
\begin{eqnarray}
\delta R = \frac{h}{2}(R + 4\Lambda).
\end{eqnarray}
Using the previous relations we find, after making the expression dimensionless by dividing by $H_0^2$, the relation
\begin{eqnarray}
\label{pv}
\delta R = - 6\Omega_m h.
\end{eqnarray}
Hence, Eq. (\ref{pe1}) becomes
\begin{eqnarray}
\label{eq1b}
\ddot h + 2H \dot h + 3\Omega_mh = 12\pi G(\delta\rho + \delta p),
\end{eqnarray}
It is important to remark that, due to relation (\ref{pv}) the scalar perturbation $h$ is directly connected with the
{\it matter component} density contrast. In this sense the relation is essentially the same as it is found in the $\Lambda$CDM model up to a numerical factor.

Now we perturb the conservation laws such that
\begin{eqnarray}
\delta({T^{\mu\nu}}_{;\mu}) + \frac{h^{\mu\nu}}{4}T_{;\mu} - \frac{g^{\mu\nu}}{4}\delta T_{;\mu} = \frac{1}{32\pi G}\biggr\{- h^{\mu\nu}R_{;\mu} + g^{\mu\nu}\delta R_{;\mu}\biggl\}.
\end{eqnarray}

This general expression leads to two perturbed equations:
\begin{eqnarray}
\label{eq2b}
\dot{\tilde\delta} + \frac{4}{3}\biggr[\theta - \biggr(1 - \frac{\Omega_m}{\Omega_\gamma}\biggl)\frac{\dot h}{2}\biggl] &=& 0,\\
\label{eq3b}
\dot\theta + H\theta &=& \frac{k^2}{a^2}\biggr\{\frac{\tilde\delta}{4}- \frac{\Omega_m}{\Omega_\gamma}h\biggl\}.
\end{eqnarray}
In these expressions,
\begin{eqnarray}
\tilde \delta = \frac{\delta\tilde\rho}{\tilde\rho},\quad \theta &=& \partial_i\delta u^i, \quad \delta\tilde\rho = \rho + p,\\
\Omega_m  = \frac{\Omega_{m0}}{a^3}\quad &,& \quad \Omega_\gamma =  \frac{\Omega_{\gamma0}}{a^4},\\
\Omega_{m0} = \frac{8\pi G}{3}\frac{\rho_{m0}}{H_0^2} \quad &,& \Omega_{\gamma0} = \frac{8\pi G}{3}\frac{\rho_{\gamma0}}{H_0^2}.
\end{eqnarray}

The final set of equations is given by,
\begin{eqnarray}
\label{eq1f}
\ddot h + 2H \dot h + 3\Omega_m h &=& 6\Omega_\gamma\tilde\delta,\\
\dot{\tilde\delta} + \frac{4}{3}\biggr[\theta - \biggr(1 - \frac{3}{4}\frac{\Omega_m}{\Omega_\gamma}\biggl)\frac{\dot h}{2}\biggl] &=& 0,\\
\dot\theta + H\theta &=& \frac{k^2}{a^2}\biggr\{\frac{\tilde\delta}{4}- \frac{3}{8}\frac{\Omega_m}{\Omega_\gamma}h\biggl\}.
\end{eqnarray}

Now we change to the scale factor as the dynamical variable. Making the equations dimensionless, we end up with the following set of equations.
\begin{eqnarray}
h''+ \biggr\{\frac{H'}{H} + \frac{3}{a}\biggl\}h' + 3\frac{\Omega_{m}}{a^2H^2}h &=& 6\frac{\Omega_\gamma}{H^2 a^2}\tilde\delta,\\
\tilde\delta' + \frac{4}{3}\biggr\{\frac{\theta}{Ha} - \biggr(1 - {3}{4}\frac{\Omega_m}{\Omega_\gamma}\biggl)\frac{h'}{2}\biggl\} &=&  0,\\
\theta' + \frac{\theta}{a} &=& \frac{k^2}{a^3H}\bigg\{\frac{\tilde\delta}{4} - \frac{3}{8}\frac{\Omega_{m}}{\Omega_\gamma}h\biggl\},
\end{eqnarray}
with
\begin{eqnarray}
H = \sqrt{\frac{\Omega_{m0}}{a^3} + \frac{\Omega_{\gamma0}}{a^4} + \Omega_{\Lambda0}}.
\end{eqnarray}

\begin{figure}[!t]
\includegraphics[width=0.45\textwidth]{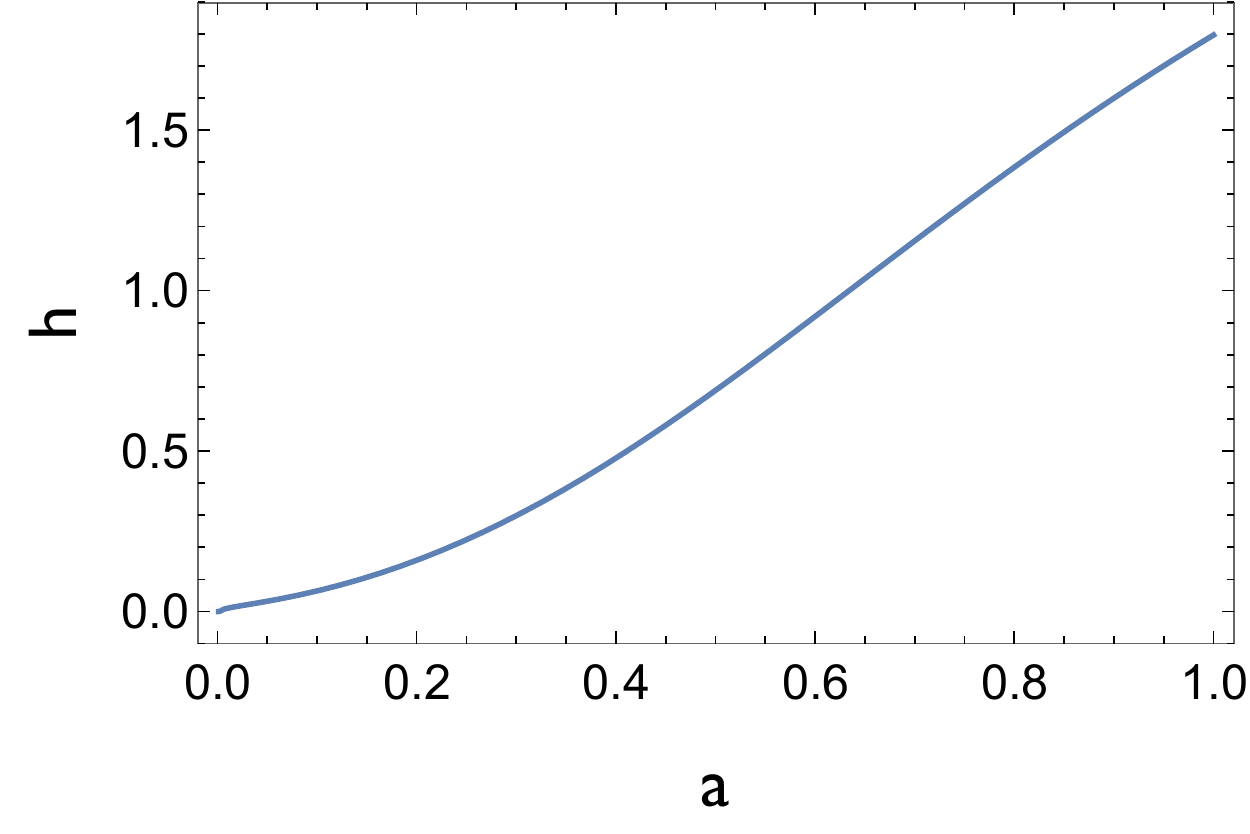}
\includegraphics[width=0.44\textwidth]{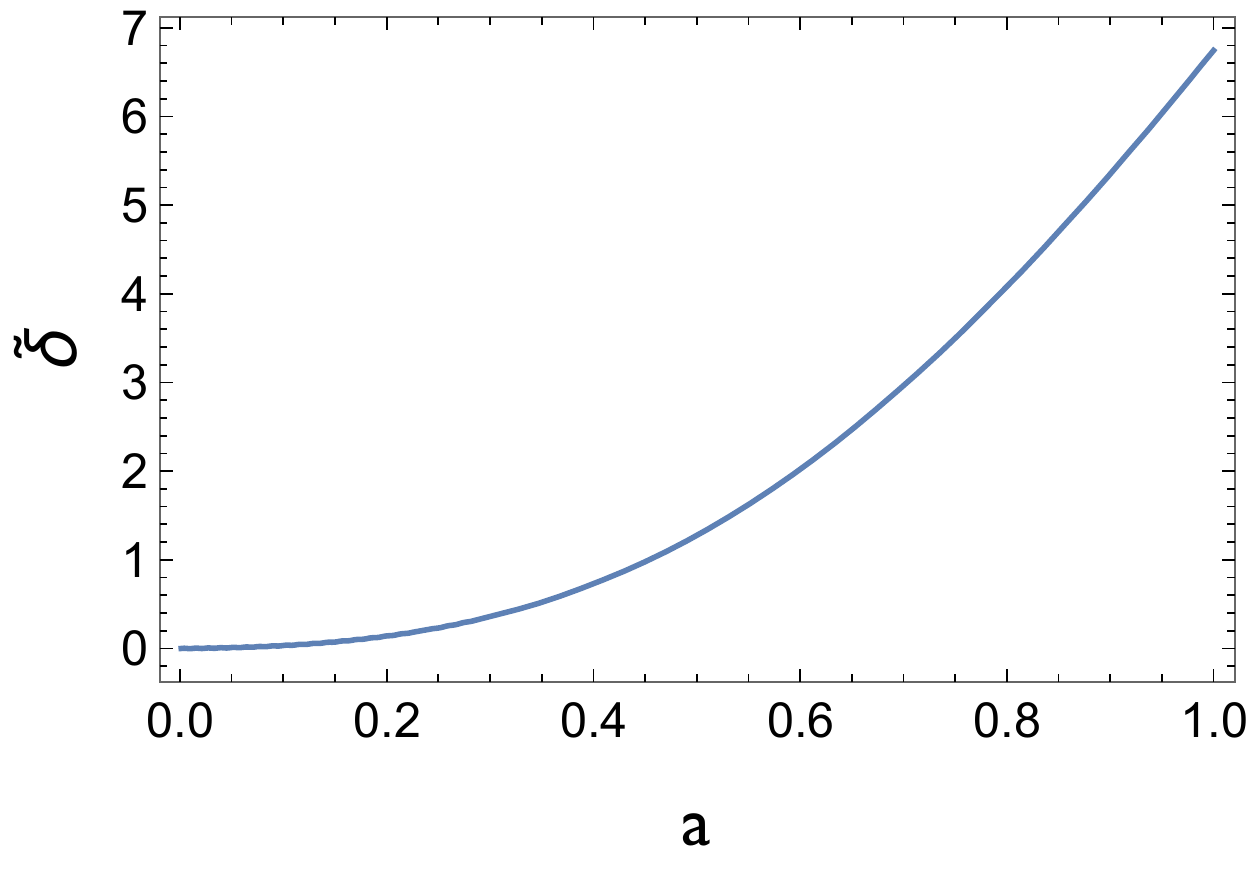}
\caption{Growth of density perturbations for $\Omega_{m0} = 0.01$ (left)  and $\Omega_{m0} = 0.3$ (right).}
\label{fig-a}
\end{figure}

Figures (\ref{fig-a}) display the behavior of the density perturbations (directly connected with $h$) for two different values of
the density parameter $\Omega_{m0}$. The scale $k = 0.1 h Mpc^{-1}$ has been used($h$ in this expression is the reduced Hubble parameter).

\begin{figure}[!t]
\includegraphics[width=0.3\textwidth]{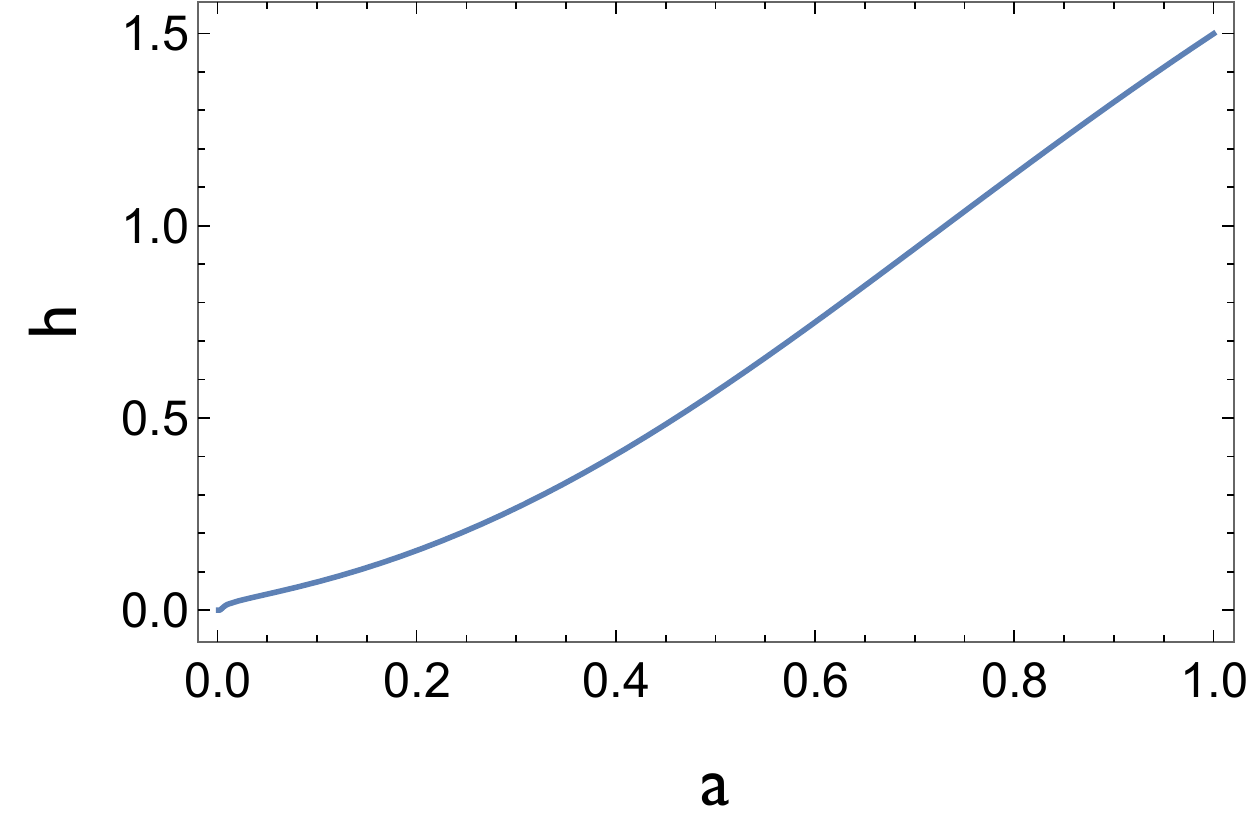}
\includegraphics[width=0.29\textwidth]{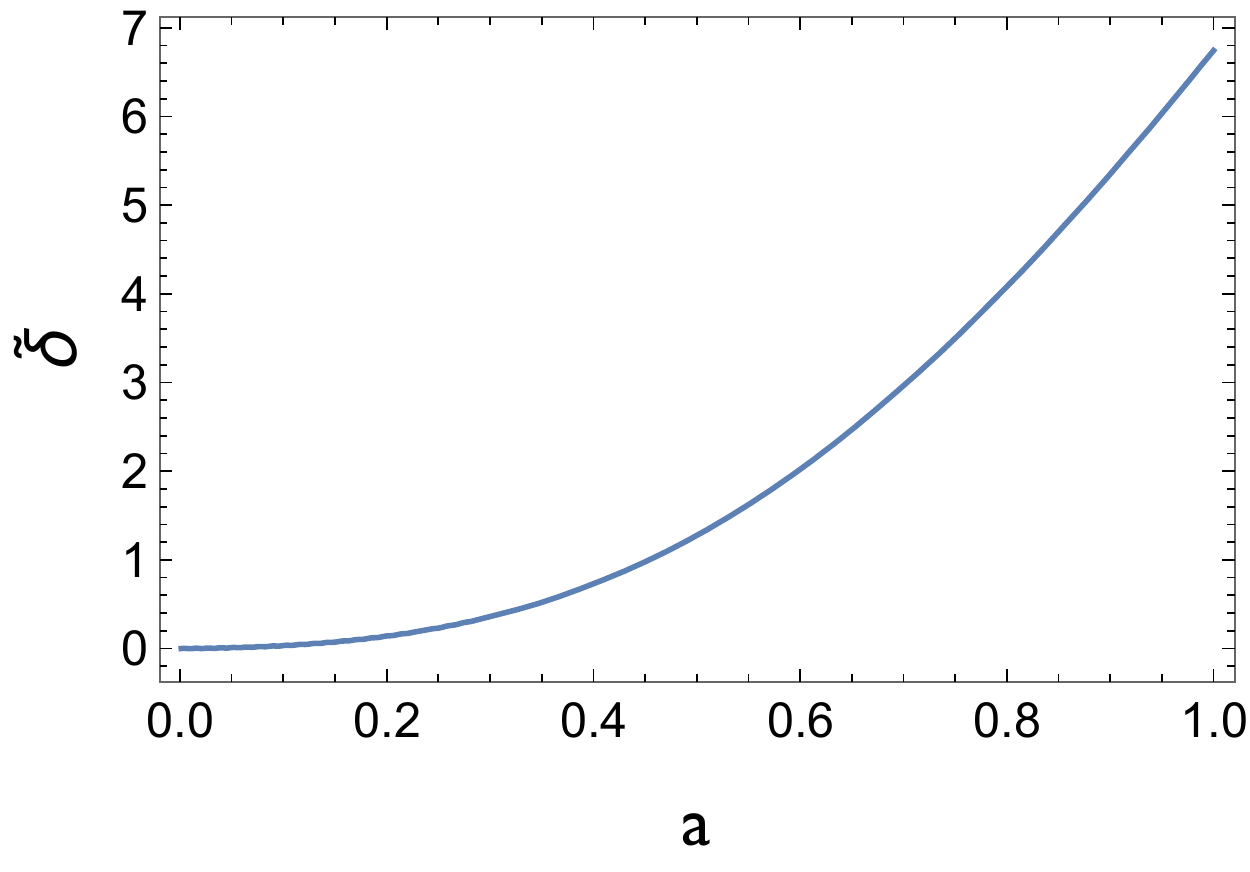}
\includegraphics[width=0.3\textwidth]{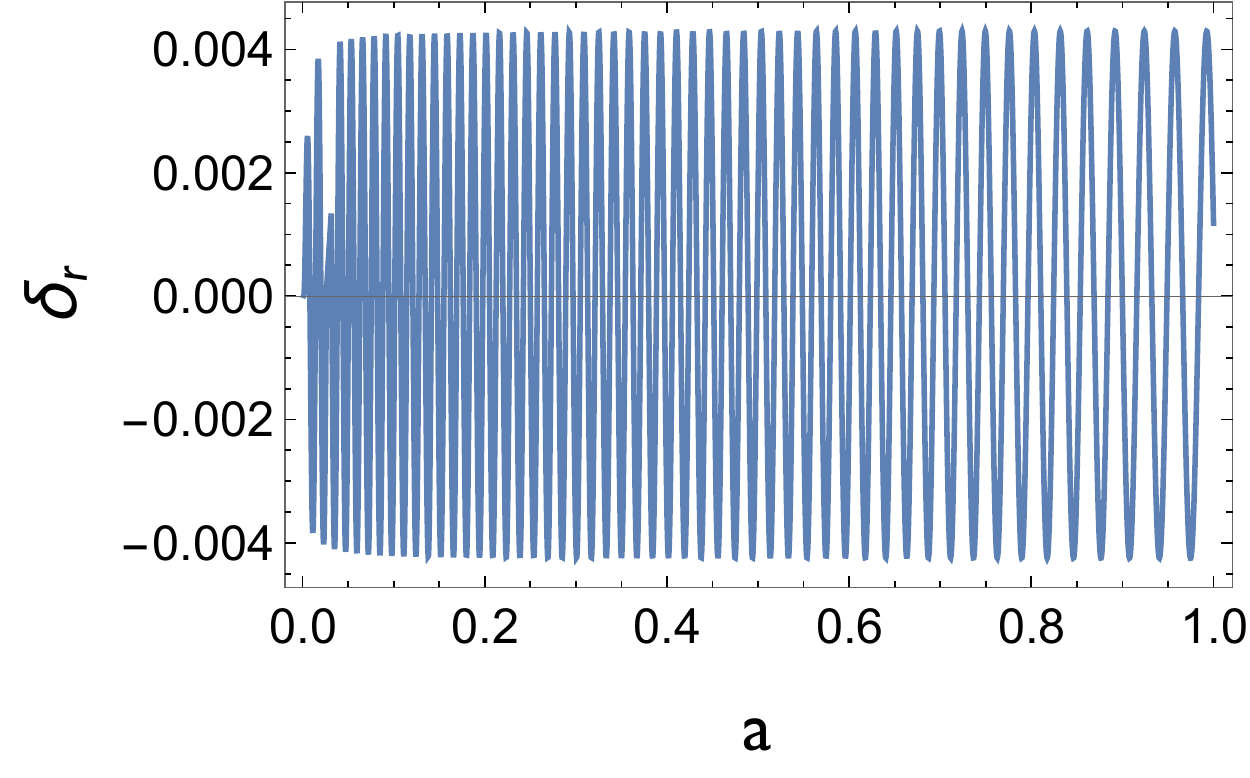}
\caption{Growth of density perturbations for $\Omega_{m0} = 0.3$ and $\Omega_r = 10^{-5}$. On the left panel the perturbation on
the pressureless component is displayed, in the center panel the perturbation on the effective radiative fluid, and in the right panel the perturbations in the true radiative fluid.}
\label{fig-b}
\end{figure}

In figures (\ref{fig-b}) the behavior of the "radiative" fluid perturbations is displayed. In fact, this fluid behaves as the usual matter perturbation with no oscillations, as in the previous section. This behavior does not depend on the equation of state for $p$. All equations are only sensible to the combination $\rho + p$. Hence, even if at background level the fluid is similar to the radiative fluid, at perturbative level on the other hand, the ansatz (\ref{R4Lambdak}) yields to behavior similar to pressureless matter. 

We can test these features by introducing a {\it true} radiative fluid $T^{\mu\nu}_r$, for which $p_r = \frac{\rho_r}{3}$, conserving separately, i.e.,  $T{^{\mu\nu}_r}_{;\mu} = 0$. In this case the perturbed equations read,
\begin{eqnarray}
\label{eq1f}
\ddot h + 2H \dot h + 3\Omega_m h &=& 6\Omega_\gamma\tilde\delta + 6\Omega_r\delta_r,\\
\dot{\tilde\delta} + \frac{4}{3}\biggr[\theta - \biggr(1 - \frac{3}{4}\frac{\Omega_m}{\Omega_\gamma}\biggl)\frac{\dot h}{2}\biggl] &=& 0,\\
\dot\theta + H\theta &=& \frac{k^2}{a^2}\biggr\{\frac{\tilde\delta}{4}- \frac{3}{8}\frac{\Omega_m}{\Omega_\gamma}h\biggl\}\\
\dot{\delta}_r + \frac{4}{3}\biggr[\theta_r - \frac{\dot h}{2}\biggl] &=& 0,\\
\dot\theta + H\theta &=& \frac{k^2}{a^2}\frac{\delta_r}{4}.
\end{eqnarray}
In these equations the subscript $r$ indicates the {\it true} radiative fluid with its density and velocity perturbations.

\begin{figure}[!t]
\includegraphics[width=0.4\textwidth]{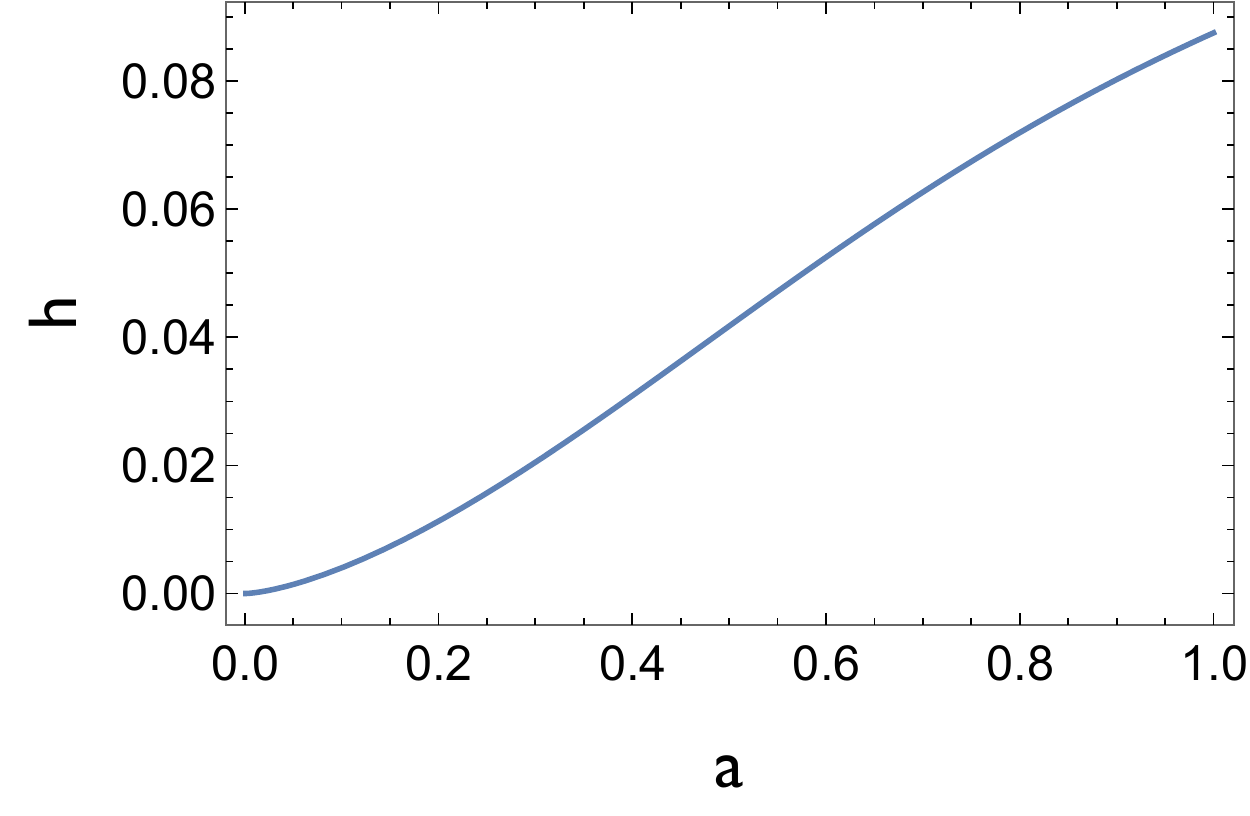}
\includegraphics[width=0.42\textwidth]{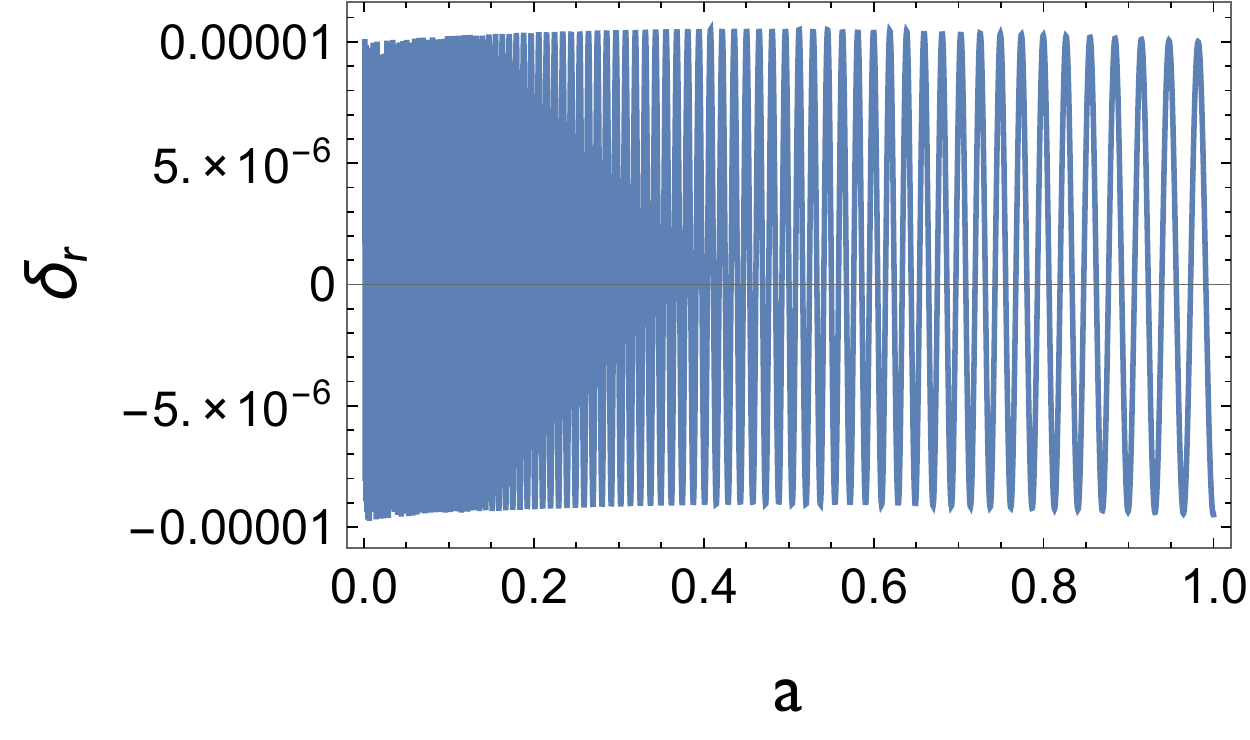}
\caption{Growth of density perturbations for $\Omega_{m0} = 0.3$ and $\Omega_r = 10^{-5}$ for the $\Lambda$CDM case. On the left panel the perturbation on
the pressureless component is displayed and in the right panel the perturbations for the radiative fluid.}
\label{fig-c}
\end{figure}

We can compare this behavior with the usual $\Lambda$CDM standard case. The perturbed equations for the $\Lambda$CDM model are

\begin{eqnarray}
\label{eq1s}
\ddot h + 2H \dot h - \frac{3\Omega_m h}{2} &=&  6\Omega_r\delta_r,\\
\dot{\delta}_r + \frac{4}{3}\biggr[\theta_r - \frac{\dot h}{2}\biggl] &=& 0,\\
\dot\theta + H\theta &=& \frac{k^2}{a^2}\frac{\delta_r}{4}.
\end{eqnarray}
The behavior for the pressureless matter (from the metric perturbation) and the perturbed radiative fluid are displayed in Fig. (\ref{fig-c}). 

The set of equation (100)-(104) has a different structure in comparison to the standard cosmology, represented by Eqs. (105)-(107), due to the appearance of the new source terms to the density contrast $\tilde{\delta}$. Indeed, such new aspects emerge from the geometrical sector of the theory and dos not have a fluid-type counterpart. An interesting qualitative result is that $\tilde{\delta}$ indeed grows in amplitude with time as expected to give rise to cosmological large scale structure. However, the use of matter growth data can be helpful to discriminate quantitatively between such models but this is beyond the aims of this work.

\section{Conclusions}

We have designed in this work a non-conservative version o unimodular theories of gravity. Even if not starting from the unimodular constraint on $\sqrt{-g}$ our main idea was to keep in mind the unimodular field equations and then to evade the imposition that the total energy-momentum tensor has to be conserved. It is worth noting that by imposing $T^{\mu\nu}_{\, ;\mu}=0$ the unimodular field equations are formally the same as GR + cosmological constant.

Since $T^{\mu\nu}_{\, ;\mu}\neq 0$ new constraints must be imposed to the field equations. In Section III we explored the trivial case given by a constant Ricci scalar. This analysis has a pedagogical appeal since we ended up with the same expansion as the one given by GR sourced with a radiative fluid and cosmological constant. At the background level the effective dynamics interpolates from a pure radiation dominated Universe to the de Sitter phase in the future time limit without having a matter-dominated epoch. 
Scalar perturbations behave similarly, at least for the case of the background solution represented by the hyperbolic sinus, to the
corresponding General Relativity case mixing radiation and a cosmological constant. Such equivalence emerges uniquely due to the ansatz employed. However, even in this case, some special configurations may appear, which are not presented in the GR counterpart case, as the possibility to introduce a conserved matter component without spoiling the background solution. 

Moreover, in order to describe the present Universe it is necessary to relax the condition $R =$ constant. It has to be violated if we want to obtain a matter dominated phase during the evolution of the Universe. We have found in Section III that the non-trivial case $\sqrt{-g}\{R + 4\Lambda\} = l$, where $l$ is constant, yields to a $\Lambda$CDM like expansion. There is new feature here since matter-like behavior has a geometric origin. The absence of such matter dominated phase in the model of Section II is surely one of the reasons responsable to rule out this approach since it is not adapted to the description of the present Universe. Then, the condition $\sqrt{-g}\{R + 4\Lambda\} = l$ leads to a transient matter dominated phase. However, the perturbative analysis becomes much more complex. This ansatz leads to important new features: the effective radiative fluid perturbatively behaves as a matter component, and in order to recover the usual radiative behavior, a traceless fluid which conserves separately must be introduced. In general $\sqrt{-g}\{R + 4\Lambda\} = l$ leads to a perturbative behavior completely different from the $\Lambda$CDM framework even if the background expansion is essentially the same. The latter reinforces the viability of the model. This unexpected effect should be further investigated in details in light of recent data. Perhaps a more natural extension is to implement holographic hypothesis similar to
that used in many analysis with interaction in the dark sector, a possibility that we intend to explore in future works.

Issues concerning other astrophysical and cosmological aspects of this theory should also be further investigated. For example, applying a time reversal symmetry ($t \rightarrow -t$) to the first model investigated in the present paper leads to a transition from the de Sitter expansion to a pure radiation behavior. Such phenomenology can also be explored in the realm of inflationary models which in principle is very attractive.

It is worth noting that Rastall-like theories represent just a prototype for nonconservative gravity. In this work we have explored the nonconservative unimodular gravity via this approach but another understanding of nonconservative gravity, which can also be further designed in relation to unimodular gravity, can be found in Refs. \cite{Lazo:2017udy, Fabris:2017msx}. We hope to explore deeply those extensions of the model presented here in future works.

\bigskip

\noindent
{\bf Acknowledgements:} We thank CNPq (Brazil) and FAPES (Brazil) for financial support. We are grateful to Davi C. Rodrigues for many useful discussions.


\begin{thebibliography}{}

\bibitem{wei} S. Weinberg, Rev. Mod. Phys. 61, 1 (1989).

\bibitem{diego} D. Saez-Gomez, Phys. Rev. {\bf D93}, 124040(2016).

\bibitem{Gao:2014nia} 
  C.~Gao, R.~H.~Brandenberger, Y.~Cai and P.~Chen,
  JCAP {\bf 1409}, 021 (2014)

\bibitem{Basak:2015swx} 
  A.~Basak, O.~Fabre and S.~Shankaranarayanan,
  Gen.\ Rel.\ Grav.\  {\bf 48}, no. 10, 123 (2016)

\bibitem{Padilla:2014yea} 
  A.~Padilla and I.~D.~Saltas,
  Eur.\ Phys.\ J.\ C {\bf 75}, no. 11, 561 (2015)

\bibitem{Nojiri:2015sfd} 
  S.~Nojiri, S.~D.~Odintsov and V.~K.~Oikonomou,
  JCAP {\bf 1605}, no. 05, 046 (2016)


\bibitem{Nojiri:2016ygo} 
  S.~Nojiri, S.~D.~Odintsov and V.~K.~Oikonomou,
  Phys.\ Rev.\ D {\bf 93}, no. 8, 084050 (2016)
  doi:10.1103/PhysRevD.93.084050
  [arXiv:1601.04112 [gr-qc]].


\bibitem{Nojiri:2016plt} 
  S.~Nojiri, S.~D.~Odintsov and V.~K.~Oikonomou,
  Mod.\ Phys.\ Lett.\ A {\bf 31}, no. 30, 1650172 (2016)
  doi:10.1142/S0217732316501728
  [arXiv:1605.00993 [gr-qc]].


\bibitem{Rajabi:2017alf} 
  F.~Rajabi and K.~Nozari,
  Phys.\ Rev.\ D {\bf 96}, no. 8, 084061 (2017)

\bibitem{Houndjo:2017jsj} 
  M.~J.~S.~Houndjo,
  Eur.\ Phys.\ J.\ C {\bf 77}, no. 9, 607 (2017)
  
  \bibitem{mukhanov}
  A.H. Chamseddine, V. Mukhanov and A. Vikman, JCAP 1{\bf 406}, 017(2014).


\bibitem{Nojiri:2016ppu} 
  S.~Nojiri, S.~D.~Odintsov and V.~K.~Oikonomou,
  Class.\ Quant.\ Grav.\  {\bf 33}, no. 12, 125017 (2016)
  doi:10.1088/0264-9381/33/12/125017
  [arXiv:1601.07057 [gr-qc]].

\bibitem{Barrow:2019gup} 
  J.~D.~Barrow and S.~Cotsakis,
  ``Inflation Without a Trace of Lambda,''
  arXiv:1907.02928 [gr-qc].


\bibitem{Rastall:1973nw} 
  P.~Rastall,
  Phys.\ Rev.\ D {\bf 6}, 3357 (1972).


\bibitem{Oliveira:2016ooo} 
  A.~M.~Oliveira, H.~E.~S.~Velten and J.~C.~Fabris,
  Phys.\ Rev.\ D {\bf 93}, no. 12, 124020 (2016)
  [arXiv:1602.08513 [gr-qc]].

\bibitem{visser} 	
M. Visser, Phys. Lett. {\bf B782}, 83 (2018).

\bibitem{darabi} 	
F. Darabi, H. Moradpour, I. Licata, Y. Heydarzade and C. Corda, Eur. Phys. J. {\bf C78}, 25(2018).

  
\bibitem{ellis}  G.F.R. Ellis, Gen. Rel. Gravit. {\bf 46}, 1619 (2014)
  
\bibitem{barrow} J.D. Barrow, S. Cotsakis, {\it Inflation without a trace of Lambda}, arXiv: 1907.0292[gr-qc].

\bibitem{Alvarez:2007nn} 
  E.~Alvarez and A.~F.~Faedo,
  Phys.\ Rev.\ D {\bf 76}, 064013 (2007)

\bibitem{Barcelo:2014mua} 
  C.~Barcel\'o, R.~Carballo-Rubio and L.~J.~Garay,
  Phys.\ Rev.\ D {\bf 89}, no. 12, 124019 (2014)

\bibitem{man} Ph. Mannheim and K. Kazanas, ApJ 342, 635(1989).


\bibitem{pad} T. Padmanabhan, Int. J. Mod. Phys. {\bf D25}, 1630020 (2016).

\bibitem{weinberg} S. Weinberg, {\bf Gravitation and Cosmology}, Wiley, New York (1972).


\bibitem{Lazo:2017udy} 
  M.~J.~Lazo, J.~Paiva, J.~T.~S.~Amaral and G.~S.~F.~Frederico,
  Phys.\ Rev.\ D {\bf 95}, no. 10, 101501 (2017)


\bibitem{Fabris:2017msx} 
  J.~C.~Fabris, H.~Velten, T.~R.~P.~Caramês, M.~J.~Lazo and G.~S.~F.~Frederico, Int. J. Mod. Phys. {\bf D27}, 1841006 (2018).


\end{thebibliography}
\end{document}